# Imaging the dynamics of free-electron Landau states

P. Schattschneider[1,2,3], Th. Schachinger[1], M. Stöger-Pollach[3], S. Löffler[3], A. Steiger-Thirsfeld[3], K.Y. Bliokh[4,5], and F. Nori[5,6]

[1]*Institute of Solid State Physics, Vienna University of Technology, Wiedner Hauptstraße 8-10, 1040 Vienna, Austria*
[2]*LMSSMat (CNRS UMR 8579) Ecole Centrale Paris, F-92295 Châtenay-Malabry, France*
[3]*University Service Centre for Electron Microscopy, Vienna University of Technology, Wiedner Hauptstraße 8-10, 1040 Vienna, Austria*
[4]*iTHES Research Group, RIKEN, Wako-shi, Saitama 351-0198, Japan*
[5]*Center for Emergent Matter Science, RIKEN, Wako-shi, Saitama 351-0198, Japan*
[6]*Physics Department, University of Michigan, Ann Arbor, Michigan 48109-1040, USA*

Landau levels and states of electrons in a magnetic field are fundamental quantum entities underlying the quantum-Hall and related effects in condensed matter physics. However, the real-space properties and observation of Landau wave functions remain elusive. Here we report the first real-space observation of Landau states and the internal rotational dynamics of *free* electrons. States with different quantum numbers are produced using nanometer-size electron vortex beams, with a radius chosen to match the waist of the Landau states, in a quasi-uniform magnetic field. Scanning the beams along the propagation direction, we reconstruct the rotational dynamics of the Landau wave functions with angular frequency ~100 GHz. We observe that Landau modes with different azimuthal quantum numbers belong to three classes, which are characterized by rotations with zero, Larmor, and cyclotron frequencies, respectively. This is in sharp contrast to the uniform cyclotron rotation of classical electrons, and in perfect agreement with recent theoretical predictions.

## Introduction

Classical electrons in a uniform magnetic field propagate freely along the field and form confined circular orbits in the plane perpendicular to the field. The angular velocity of such orbiting is constant and known as the cyclotron frequency. Accordingly, quantum-mechanical eigenstates of a scalar electron in a uniform magnetic field are localized in the transverse plane, and are characterized by *two* quantum numbers. Excluding the longitudinal motion of the electron (as, e.g., in 2D condensed-matter systems), this leads to the quantization of the energy levels, which are degenerate and are characterized by a *single* quantum number. Quantum electron states and their corresponding energy levels in a magnetic field were described by Fock [1], Landau [2], and Darwin [3] in the early days of quantum theory, and are commonly referred to as Landau states and Landau levels.

Landau eigenstates play a key role in various solid-state phenomena, such as the diamagnetism of metals, as well as quantum-Hall, Shubnikov–De Haas, and De Haas–van Alphen effects [4–6]. Landau energy levels reveal themselves in quantum-Hall conductance plateaus [6], they are measured spectroscopically [7], and recently they attracted enormous attention in relation to graphene systems [8–12]. However, Landau levels are highly degenerate and do not provide information about the actual state and spatial distribution of the electron. Moreover, the drift of the states in an external random potential blurs the picture in condensed-matter systems [13,14], and it is impossible to observe the fast rotational dynamics of electrons in such systems. Although considerable progress was achieved recently in Fourier-analysis of Landau modes [15] (based on a *single radial* quantum number), their real-space properties remain elusive. Thus, the observation of spatially-resolved Landau eigenstates and their internal dynamics remains a challenging problem.

Landau states can appear not only in condensed-matter systems, but also for *free* electrons in a uniform magnetic field. Recently, we argued [16] that, allowing free propagation along the magnetic field, the Landau states represent non-diffracting versions of the so-called *electron vortex beams* [17–20]. Furthermore, both the *radial* and *azimuthal* (vortex) quantum numbers of Landau modes crucially determine their properties and evolution. Electron vortex beams were predicted [17] and generated in transmission electron microscopes [18–24] few years ago, and they promise applications in various areas of both fundamental and applied physics [24–34] (for a review, see [35]). Here we report the real-space observation of individual Landau eigenstates, which are formed by free-electron vortex beams in a uniform magnetic field inside a transmission electron microscope. We measure the fast rotational dynamics of electrons within different states, and reveal their unusual non-classical behaviour. Instead of cyclotron orbiting, we observe that Landau electrons rotate with *three different angular velocities*, determined by the vortex quantum number.

## Results

**Rotational dynamics of electrons in quantum Landau states.** The Landau states of an electron in a $z$-directed homogeneous magnetic field $B>0$ and uniform gauge with azimuthal vector-potential $A_\varphi = Br/2$ are described by cylindrical vortex wave functions [1,3,16]

$$\psi_{m,n} \propto \left(\frac{r}{w_B}\right)^{|m|} L_n^{|m|}\left(\frac{r^2}{w_B^2}\right) \exp\left(-\frac{r^2}{2w_B^2}\right) \exp\left[i(m\varphi + k_z z)\right]. \tag{1}$$

Here $(r,\varphi,z)$ are cylindrical coordinates, $m = 0, \pm 1, \pm 2, ...$ and $n = 0,1,2,...$ are the azimuthal and radial quantum numbers, respectively, $w_B = \sqrt{2\hbar/|e|B}$ is the magnetic length parameter, $L_n^{|m|}$ are the generalized Laguerre polynomials, and $k_z$ is the wave number of the free longitudinal electron motion. The transverse energy of the electron is quantized according to Landau levels [1,3,16]

$$E_\perp = \hbar\Omega(2N+1), \quad \text{where} \quad N = n + (|m| + m)/2 = 0,1,2,... \tag{2}$$

is the principal Landau quantum number and $\Omega = |e|B/2\mu_e$ is the Larmor frequency corresponding to the electron charge $e = -|e|$ and mass $\mu_e$. Although the classical electron dynamics is determined by the cyclotron frequency $\omega_c = 2\Omega$, it is the Larmor frequency that is fundamental in the quantum evolution of electrons [16,36].

The Landau states (1) resemble $z$-propagating free-space vortex beams [17–20], which are characterized by the phase factor $\exp(im\varphi)$, circulating $m$-dependent azimuthal current, and kinetic orbital angular momentum $L_z = \hbar m$ per electron. However, the kinetic orbital angular momentum of the Landau states differs significantly from that of free-space beams and is determined by the principal quantum number: $L_z = E_\perp / \Omega = \hbar(2N+1) > 0$ [16]. It is this positive angular momentum and the corresponding negative magnetic moment $M_z = (e/2\mu_e)L_z < 0$ that are responsible for the diamagnetism of electrons predicted by Landau [2,3]. The nontrivial angular momentum of the Landau states appears because the gauge-invariant probability current is modified by the presence of the vector potential **A** [37]: $\mathbf{j} = \mu_e^{-1}\left[\hbar\,\text{Im}(\psi^*\nabla\psi) - e\mathbf{A}|\psi|^2\right]$. According to this, the current in the Landau states (1) becomes [16]

$$\mathbf{j} = \frac{\hbar}{\mu_e}\left[\frac{1}{r}\left(m + \frac{r^2}{w_B^2}\right)\overline{\varphi} + k_z \overline{\mathbf{z}}\right]|\psi|^2, \qquad (3)$$

where $\overline{\varphi}$ and $\overline{\mathbf{z}}$ are the unit vectors of the corresponding coordinate axes. Here the $m$-dependent azimuthal term originates from the free-space vortex current, whereas the second azimuthal term describes the contribution from the vector-potential $A_\varphi = Br/2$.

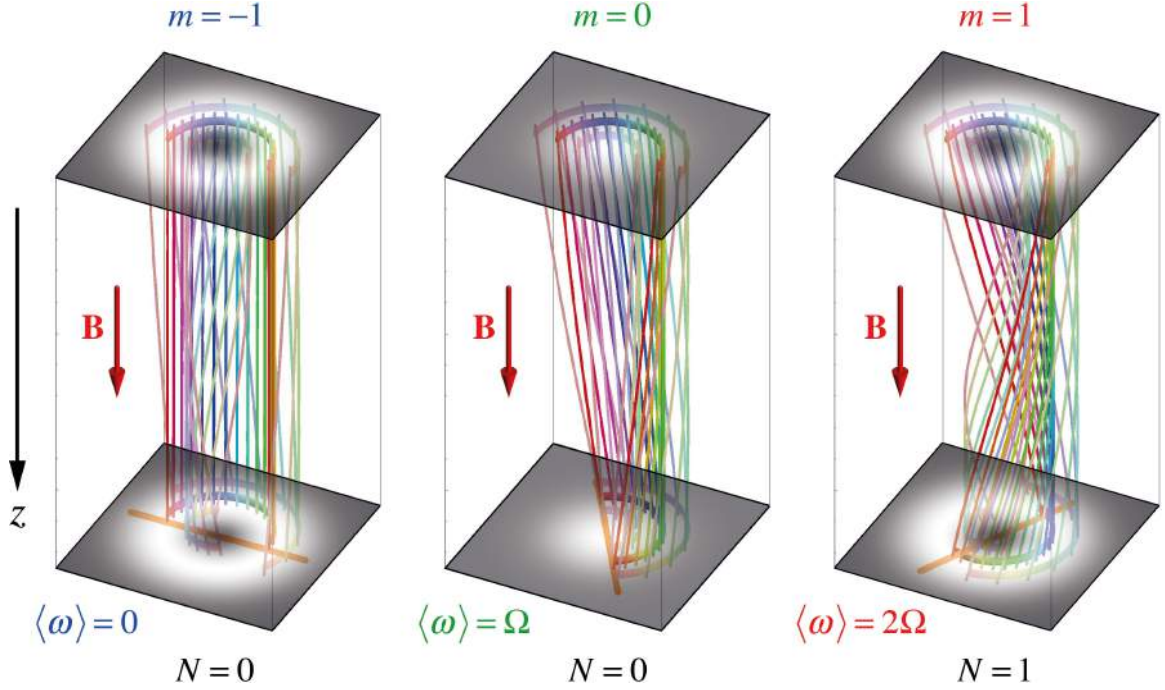

**Figure 1. Landau states with different azimuthal indices and electron Bohmian trajectories.** Gray-scale plots show the transverse probability distributions $|\psi|^2$ of the states (1) with radial quantum number $n = 0$ and azimuthal numbers $m = -1, 0, 1$ (which correspond to the Landau levels (2) with $N = 0, 0, 1$). Three-dimensional streamlines of the probability current in equation (3) (i.e., Bohmian trajectories) inside these states are shown for three different radii $r/w_B = 0.6, 1, 1.4$, in the truncated azimuthal range $\varphi|_{z=0} \in (0, \pi)$, and for the Larmor propagation distance $z \in (0, v/\Omega)$ ($v$ is the electron velocity). Different trajectories are marked by different colours for the sake of convenience. The spiralling of the Bohmian trajectories at the maximal-intensity radii $r = w_B\sqrt{|m|}$ (indicated by the orange lines in the bottom planes) corresponds to the $m$-dependent internal rotational dynamics with frequencies $\langle\omega\rangle$ of Eq. (4).

Equation (3) describes the spiralling of the Landau electron about the magnetic-field direction in the sense of *Bohmian trajectories*, i.e., streamlines of the probability current [38–41]. The expectation value of the electron's angular velocity, $\omega(r) = v_\varphi(r)/r$ (where $\mathbf{v} = \mathbf{j}/|\psi|^2$ is the local Bohmian velocity) can be obtained from Eqs. (1) and (3):

$$\langle\omega\rangle = \begin{cases} 0 & \text{for } m < 0 \\ \Omega & \text{for } m = 0 \\ 2\Omega & \text{for } m > 0 \end{cases}, \qquad (4)$$

where we used $\langle r^{-2} \rangle \equiv \int |\psi_{m,n}|^2 r^{-1} dr / \int |\psi_{m,n}|^2 r\, dr = w_B^{-2}/|m|$, for $m \neq 0$. Equation (4) represents a highly surprising result. It shows that the rotation of electrons in a magnetic field in the quantum picture is drastically different from the uniform classical orbiting. Instead of rotation with a single cyclotron frequency $\omega_c = 2\Omega$, states with negative, zero, and positive azimuthal indices rotate with zero, Larmor, and cyclotron rates, respectively. This result is independent of the radial index $n$, and is also valid for any superposition of modes with different $n$ and the same $m$ as long as $\langle r^{-2} \rangle = w_B^{-2}/|m|$. The probability-density distributions and internal Bohmian trajectories in the Landau states (1) with $n = 0$ and $m = -1, 0, 1$ (which correspond to the degenerate Landau levels (2) with $N = 0, 0, 1$) are shown in Fig. 1. One can see that in the $m = 0$ mode the trajectory rotation is uniform and corresponds to the Larmor frequency $\Omega$. At the same time, the rotations inside the $m \neq 0$ modes depend on the radius $r$ and coincide with the averaged values $\langle \omega \rangle$, Eq. (4), at the maximal-intensity radii $r = w_B\sqrt{|m|}$. As was recently demonstrated for photons [40,41], Bohmian trajectories can be measured experimentally using statistical averaging over many identical single-particle events without inter-particle interactions. The same conditions are realized in electron-optical measurements in electron microscopy [42].

**Experimental measurements.** The main goal of our experiment is twofold: first, the creation of free-electron Landau states, equation (1), and, second, the observation of their extraordinary internal rotational dynamics, equation (4). To produce the free-electron Landau states, we use the fact that they represent non-diffracting versions of the free-space Laguerre–Gaussian vortex beams [16,17]. Such electron vortex beams are generated in a transmission electron microscope using a holographic fork mask (a diffraction grating with a dislocation) [19,20]. The mask shown in Fig. 2a has a bar/slit ratio of 1, and it produces beams with different azimuthal indices $m = 0, \pm 1, \pm 3, \pm 5, ...$ for different diffraction orders [19]. (For other bar/slit ratios, even values of $m$ can also be produced [20].) The vortex beams are then focused with a magnetic lens, which has a region of a quasi-uniform $z$-directed strong magnetic field. Figure 2 shows a schematic diagram of our experimental setup with the converging vortex beams. We tune the parameters of the system, such that in the $z$-region of a quasi-uniform magnetic field the beam radius $w(z)$ comes close to the magnetic radius $w_B\sqrt{|m|}$ (Fig. 2b). Thus, the vortex beams approximate Landau states (1) with $n = 0$ and different azimuthal indices $m$. Then, the observation of the peculiar rotational dynamics described by equation (4) would verify that the beams indeed acquire properties of the Landau modes. The more the beam radius deviates from $w_B\sqrt{|m|}$, the more $\langle r^{-2} \rangle$ deviates from $w_B^{-2}/|m|$, changing the electron angular velocity given by equation (4).

To observe the internal rotational dynamics of equation (4) and spiralling Bohmian trajectories (Fig. 1) inside the cylindrically-symmetric beams, we borrow a technique successfully employed in optics [43–45] and also recently demonstrated for electrons [46]. Namely, we obstruct half of the beam with an opaque knife-edge stop and trace the spatial rotation of the visible part of the beam when the knife-edge is moved along the $z$ axis, Fig. 2a. Although such truncation of the beam breaks the cylindrical symmetry of the initial vortex state, it does not perturb significantly the probability currents in the visible part of the beam, so that the truncated beam approximately follows the internal Bohmian trajectories of the initial cylindrical state. Note that in condensed-matter systems the Larmor rotation of electrons is fast and cannot be observed as compared to the slow motion of the centre of mass in an external potential [13–15]. In contrast, for paraxial electrons in a TEM, the transverse Larmor dynamics is slow as compared to the relativistic longitudinal velocity of electrons: $\Omega w_B \ll v \sim c$. This allows mapping the internal dynamics onto the $z$ axis with extremely high resolution (in our experiment, the Larmor time scale $\Omega^{-1} \simeq 8\,\text{ps}$ corresponded to the propagation distance $z_L = v/\Omega \simeq 1.7\,\text{mm}$ [16]).

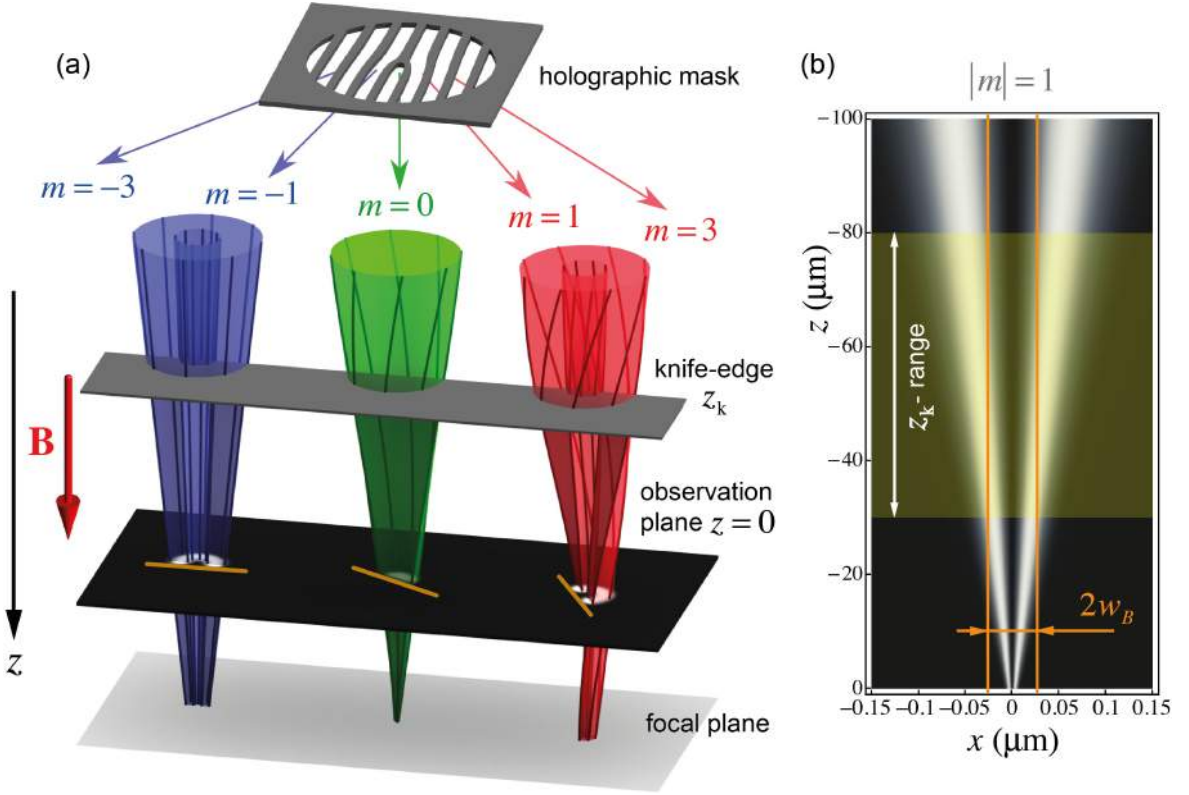

**Figure 2**. **Schematics of the experiment and the beam parameters.** **(a)** A holographic fork mask generates a row of vortex beams with different azimuthal indices $m = ..., -5, -3, -1, 0, 1, 3, 5, ...$ [19,20]. These beams are focused by a magnetic lens and are studied in the region of maximal quasi-uniform magnetic field. The focal plane is shifted few Rayleigh ranges below the observation plane $z = 0$, to reduce the Gouy-phase rotation [46,47]. A knife-edge stop is placed at $z_k < 0$, where it blocks half of each of the beams. Varying the position $z_k$ of the knife-edge, we observe spatial rotational dynamics of the cut beams propagating to the observation plane (see Fig. 1). **(b)** Intensity distribution in the $|m| = 1$ beams. The radius $w$ of the focused beams varies slowly with $z$. In the highlighted range $z \in (-80, -30)\,\mu\text{m}$, the beam radius approaches the magnetic radius, $w(z) \simeq w_B$, and the beams acquire the Landau-state properties (see Figs. 3 and 4).

The experiment was performed according to the above approach in a FEI TECNAI F20 transmission electron microscope (TEM) at $200\,\text{kV}$ acceleration voltage ($v \simeq 0.7c$). The focusing lens produced the maximal longitudinal field $B \simeq 1.9\,\text{T}$, which corresponds to the Larmor frequency $\Omega \simeq 120\,\text{GHz}$ (using the relativistic mass $\mu_e = \gamma \mu_{e0}$) and the magnetic radius $w_B \simeq 26\,\text{nm}$. Choosing $z = 0$ as the observation plane, the region of interest was $z \in (-80, -30)$ μm (Fig. 2). In this region the magnetic field was uniform up to negligible variations $\sim 10^{-2} B$ in the longitudinal component and $\sim 10^{-6} B$ in the radial component. The knife-edge was made from a Si crystal. Its position $z_k$ was varied in the region of interest (i.e., the propagation distance to the observation plane was varied by $\Delta z \simeq 50\,\mu\text{m}$) to measure the rotations of the images of the cut beam (the corresponding Larmor angle is $\Delta\varphi = \Delta z / z_L \simeq 30\,\text{mrad}$, see Fig. 3). Note that the focal plane of the beams was set at $z \simeq 8\,\mu\text{m}$, i.e., few Rayleigh ranges below the observation plane, in order to reduce the diffractive Gouy-phase rotation of the images [46,47] and to improve the accuracy of the measurements using sufficiently large vortex radii (Fig. 2a).

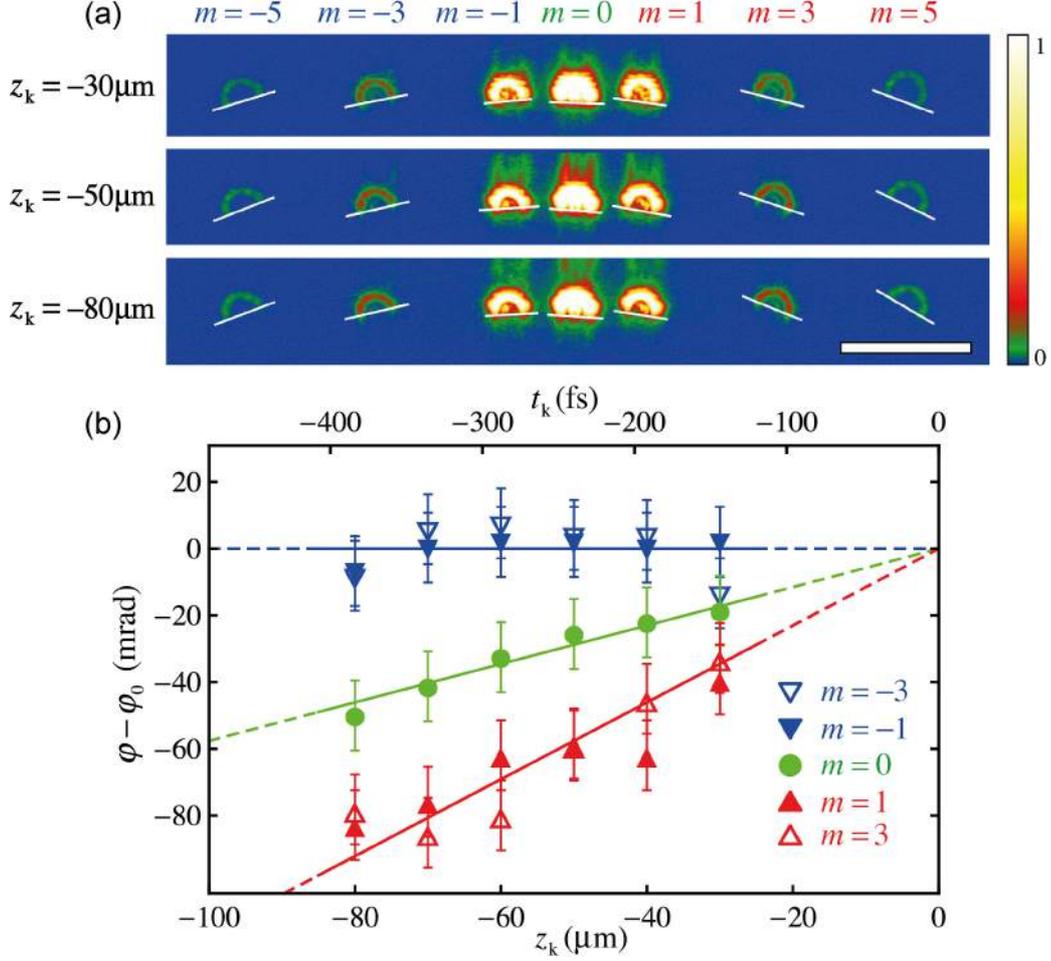

**Figure 3. Experimental images of the cut vortex beams and their $m$-dependent rotations with the propagation distance. (a)** Experimental images of the row of the cut vortex beams with different $m$ at different positions of the knife-edge, $z_k$ (Fig. 2). The scalebar is 50 nm. The opposite inclination of the opposite-$m$ states is due to the residual Gouy-phase diffraction effect [46,47]. At the same time, one can see a slow rotation of the $m>0$ modes with $z_k$, while the $m<0$ states remain motionless. A quantitative analysis of these $m$-dependent rotations is depicted in **(b)**. The azimuthal orientations of the cut modes $\varphi$ (with respect to the extrapolated reference azimuth $\varphi|_{z_k=0} = \varphi_0$), are plotted versus $z_k$ and the corresponding time scale $t_k = z_k/v$ (on the top). Three lines correspond to the zero, Larmor, and cyclotron rotations predicted for the Landau states in Eq. (4). Error bars include the uncertainty in reading, knife-edge roughness, and stage positioning.

Figures 3 and 4 show the results of the experimental measurements of the cut vortex modes at different positions $z_k$ of the knife-edge. In Fig. 3a the images of the modes with $m = -5, -3, -1, 0, 1, 3, 5$ are shown for three values of $z_k$. Note that the cut edges of the beams with opposite $m$ have opposite inclination with respect to the line joining the beams. This is the residual Gouy-phase rotation [46,47] visible at the observation plane. At the same time, a slow rotation of the $m>0$ states as a function of $z_k$ can be detected visually, while the $m<0$ modes do not experience visible rotation. A quantitative analysis of the differential rotations of modes with different $m$ is depicted in Fig. 3b. One can clearly see three different rates of rotations for modes with $m<0$, $m=0$, and $m>0$, in precise agreement with the prediction of Eq. (4) and in sharp contrast to the classical cyclotron orbiting. This confirms that electron vortices form Landau states and acquire their peculiar properties in the region of interest. We repeated measurements of the mode rotations with $z_k$ with slightly different defocus values and holographic masks (including those producing $m = \pm 2$ modes). For each of such experiments we determined the average value of the rotational velocities $\langle \omega \rangle = v \langle d\varphi / dz_k \rangle$ for different $m$. The results are shown in Fig. 4. One can clearly see that rotational frequencies in different measurements fluctuate around the theoretical values of equation (4), and their averages over all measurements are in very good agreements with the theoretical Landau-state behaviour. Note that variations of the rotational frequencies in the $m \neq 0$ modes and robustness of such frequencies for the $m = 0$ mode can be related to the radial dependence (independence) of the rotation in the $m \neq 0$ ($m = 0$) Landau states (see Fig. 1).

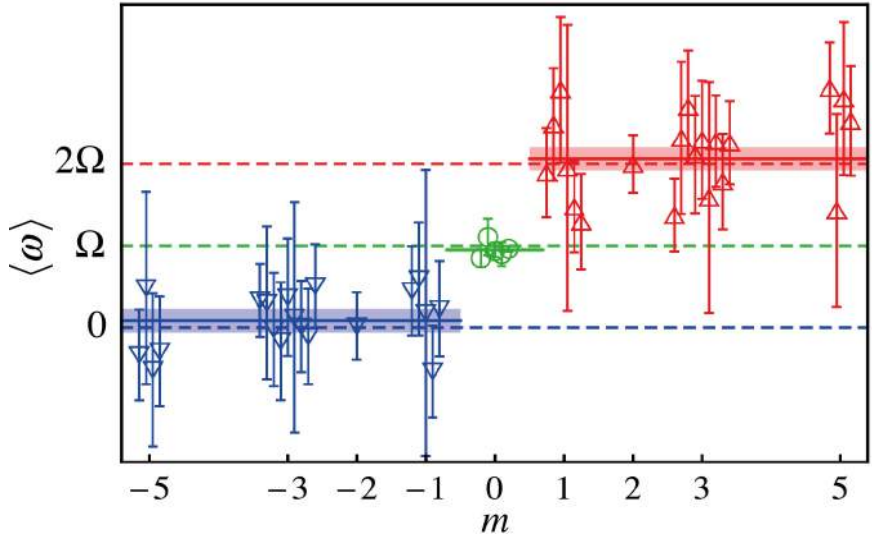

**Figure 4**. **Averaged rotational frequencies for modes with different azimuthal indices $m$.** Averaged rotational rates $\langle \omega \rangle = v \langle d\varphi / dz_k \rangle$ (such as average slopes of the data in Fig. 3b) are shown for different topological charges $m$. Different data points for the same $m$ correspond to different series of measurements, and error bars indicate the standard error of the mean in each series. The solid lines represent frequencies averaged over all measurements, while the dashed lines indicate the theoretical values predicted in Eq. (4). The average values and standard error of the mean (indicated as shaded bars) from all measurements are $\langle \omega \rangle = (0.09 \pm 0.15)\Omega$ for $m < 0$, $\langle \omega \rangle = (0.95 \pm 0.03)\Omega$ for $m = 0$, and $\langle \omega \rangle = (2.06 \pm 0.14)\Omega$ for $m > 0$. This verifies the extraordinary rotational dynamics of electrons in Landau states, which exhibit zero, Larmor, and cyclotron frequencies for the modes with $m < 0$, $m = 0$, and $m > 0$, respectively.

## Discussion

To summarize, the extraordinary $m$-dependent rotational dynamics, which is impossible in classical electron propagation, reveals the peculiar behaviour of quantum Landau states. Although it is commonly believed that the cyclotron rotation of electrons underpins Landau states, we have shown that electrons can also rotate in the quantum Bohmian picture with either zero or Larmor frequencies. (It is worth remarking that the observed $m$-dependent rotation of electrons can be related to the Aharonov–Bohm effect [48]. Indeed, it is the asymmetry of the azimuthal currents in the $m > 0$ and $m < 0$ modes, caused by the presence of the vector-potential, that is responsible for the Aharonov–Bohm phenomenon [16,48].) We emphasize several striking features and the fundamental importance of these results. First, we demonstrated the appearance of quantum Landau states within *free*-electron optics, rather than in a condensed-matter system. Second, in contrast to condensed-matter experiments and analyses, we separated modes with different *azimuthal* quantum numbers $m$ (some of which correspond to the same Landau levels) and showed that this index is crucially important for the electron rotation in a magnetic field. Third, while the observation of the fast Larmor dynamics is currently impossible in condensed-matter systems, this rotation becomes measurable for free relativistic electrons. In our setup, this allowed the detection of rotations with frequency $\Omega \sim 100$ GHz, corresponding to the energy difference $\hbar\Omega \sim 100\,\mu\text{eV}$. Thus, our results provide new insights into the fundamental properties of Landau states and pave the way towards detailed investigations of their otherwise hidden characteristics.

# References

1. Fock, V. Bemerkung zur Quantelung des harmonischen Oszillators im Magnetfeld. *Z. Physik* **47**, 446–448 (1928).
2. Landau, L. Diamagnetismus der Metalle. *Z. Physik* **64**, 629–637 (1930).
3. Darwin, C.G. The diamagnetism of free electron. *Math. Proc. Cambridge Phil. Soc.* **27**, 86–90 (1931).
4. Kittel, C. *Quantum Theory of Solids* (John Willey & Sons, 1987).
5. Marder, M.P. *Condensed Matter Physics* (John Willey & Sons, 2010).
6. Yoshioka, D. *The Quantum Hall Effect* (Springer, Berlin, 2002).
7. Main, P.C., Thornton, A.S.G., Hill, R.J.A., Stoddart, S.T., Ihn, T., Eaves, L., Benedict, K.A. & Henini, M. Landau-level spectroscopy of a two-dimensional electron system by tunneling through a quantum dot. *Phys. Rev. Lett.* **84**, 729–732 (2000).
8. Novoselov, K.S., Geim, A.K., Morozov, S.V., Jiang, D., Katsnelson, M.I., Grigorieva, I.V., Dubonos, S.V. & Firsov, A.A. Two-dimensional gas of massless Dirac fermions in graphene. *Nature* **438**, 197–200 (2005).
9. Zhang, Y., Tan, Y.-W., Stormer, H.L. & Kim, P. Experimental observation of the quantum Hall effect and Berry's phase in graphene. *Nature* **438**, 201–204 (2005).
10. Sadowski, M.L., Martinez, G., Potemski, M., Berger, C. & de Heer, W.A. Landau level spectroscopy of ultrathin graphite layers. *Phys. Rev. Lett.* **97**, 266405 (2006).
11. Li G. & Andrei, E.Y. Observation of Landau levels of Dirac fermions in graphite. *Nature Phys.* **3**, 623–627 (2007).
12. Castro Neto, A.H., Guinea, F., Peres, N.M.R., Novoselov, K.S. & Geim, A.K. The electron properties of graphene. *Rev. Mod. Phys.* **81**, 109–162 (2009).
13. Hashimoto, K., Sohrmann, C., Wiebe, J., Inaoka, T., Meier, F., Hirayama, Y., Römer, R.A., Wiesendanger, R. & Morgenstern, M. Quantum Hall Transition in Real Space: From Localized to Extended States. *Phys. Rev. Lett.* **101**, 256802 (2008).
14. Miller, D.L., Kubista, K.D., Rutter, G.M., Ruan, M., de Heer, W.A., Kindermann, M., First, P.N. & Stroscio, J.A. Real-space mapping of magnetically quantized graphene states. *Nature Phys.* **6**, 811–817 (2010).
15. Hashimoto, K., Champel, T., Sohrmann, C., Wiebe, J., Hirayama, Y., Römer, R.A., Wiesendanger, R. & Morgenstern, M. Robust nodal structure of Landau level wave functions revealed by Fourier transform scanning tunnelling spectroscopy. *Phys. Rev. Lett.* **109**, 116805 (2012).
16. Bliokh, K.Y., Schattschneider, P., Verbeeck, J. & Nori, F. Electron vortex beams in a magnetic field: A new twist on Landau levels and Aharonov–Bohm states. *Phys. Rev. X* **2**, 041011 (2012).
17. Bliokh, K.Y., Bliokh, Y.P., Savel'ev, S. & Nori, F. Semiclassical dynamics of electron wave packet states with phase vortices. *Phys. Rev. Lett.* **99**, 190404 (2007).
18. Uchida M. & Tonomura, A. Generation of electron beams carrying orbital angular momentum. *Nature* **464**, 737–739 (2010).
19. Verbeek, J., Tian, H. & Schattschneider, P. Production and application of electron vortex beams. *Nature* **467**, 301–304 (2010).

20. McMorran, B.J., Agrawal, A., Anderson, I.M., Herzing, A.A., Lezec, H.J., McClelland, J.J. & Unguris, J. Electron vortex beams with high quanta of orbital angular momenta. *Science* **331**, 192–195 (2011).
21. Schattschneider, P., Stöger-Pollach, M. & Verbeeck, J. Novel vortex generator and mode converter for electron beams. *Phys. Rev. Lett.* **109**, 084801 (2012).
22. Clark, L., Béché, A., Guzzinatti, G., Lubk, A., Mazilu, M., Van Boxem, R. & Verbeeck, J. Exploiting lens aberrations to create electron-vortex beams. *Phys. Rev. Lett.* **111**, 064801 (2013).
23. Grillo, V., Karimi, E., Gazzadi, G.C., Frabboni, S., Dennis, M.R. & Boyd, R.W. Generation of nondiffracting electron Bessel beams. *Phys. Rev. X* **4**, 011013 (2014).
24. Béché, A., Van Boxem, R., Van Tendeloo, G. & Verbeeck, J. Magnetic monopole field exposed by electrons. *Nature Phys.* **10**, 26–29 (2014).
25. Idrobo, J.C. & Pennycook, S.J. Vortex beams for atomic resolution dichroism. *J. Electron Microscopy* **60**, 295–300 (2011).
26. Verbeeck, J., Schattschneider, P., Lazar, S., Stöger-Pollach, M., Löffler, S., Steiger-Thirsfeld, A. & Van Tendeloo, G. Atomic scale electron vortices for nanoresearch. *Appl. Phys. Lett.* **99**, 203109 (2011).
27. Ivanov, I.P. Colliding particles carrying nonzero orbital angular momentum. *Phys. Rev. D* **83**, 093001 (2011).
28. Bliokh, K.Y., Dennis, M.R. & Nori, F. Relativistic electron vortex beams: angular momentum and spin-orbit interaction. *Phys. Rev. Lett.* **107**, 174802 (2011).
29. Karimi, E., Marrucci, L., Grillo, V. & Santamato, E. Spin-to-orbital angular momentum conversion and spin-polarization filtering in electron beams. *Phys. Rev. Lett.* **108**, 044801 (2012).
30. Xin, H.L. & Zheng, H. On-column 2p bound state with topological charge ±1 excited by an atomic-size vortex beam in an aberration-corrected scanning transmission electron microscope. *Microsc. Microanal.* **18**, 711–719 (2012).
31. Schattschneider, P., Schaffer, B., Ennen, I. & Verbeeck, J. Mapping spin-polarized transitions with atomic resolution. *Phys. Rev. B* **85**, 134422 (2012).
32. Mohammadi, Z., Van Vlack, C.P., Hughes, S., Bornemann, J. & Gordon, R. Vortex electron energy loss spectroscopy for near-field mapping of magnetic plasmons. *Opt. Express* **20**, 15024–15034 (2012).
33. Greenshields, C., Stamps, R.L. & Franke-Arnold, S. Vacuum Faraday effect for electrons. *New J. Phys.* **14**, 103040 (2012).
34. Ivanov, I.P. & Karlovets, D.V. Detecting transition radiation from a magnetic moment. *Phys. Rev. Lett.* **110**, 264801 (2013).
35. Verbeeck, J., Guzzinati, G., Clark, L., Juchtmans, R., Van Boxem, R., Tian, H., Beche, A., Lubk, A. & Van Tendeloo, G. Shaping electron beams for the generation of innovative measurements in the (S)TEM. *Compt. Rend. Phys.* **15**, 190–199 (2014).
36. Brillouin, L. A theorem of Larmor and its importance for electrons in magnetic fields. *Phys. Rev.* **67**, 260–266 (1945).
37. Landau L.D. & Lifshitz, E.M. *Quantum Mechanis: Non-relativistic Theory* (Butterworth-Heinemann, 1981).
38. Bohm, D., Hiley, B.J. & Kaloyerou, P.N. An ontological basis for the quantum theory. *Phys. Rep.* **144**, 321–375 (1987).
39. Wiseman, H.M. Grounding Bohmian mechanics in weak values and bayesianism. *New J. Phys.* **9**, 165 (2007).
40. Kocsis, S., Braverman, B., Ravets, S., Stevens, M.J., Mirin, R.P., Shalm, L.K. & Steinberg, A.M. Observing the average trajectories of single photons in a two-slit interferometer. *Science* **332**, 1170–1173 (2011).

41. Bliokh, K.Y., Bekshaev, A.Y., Kofman, A.G. & Nori, F. Photon trajectories, anomalous velocities, and weak measurements: a classical interpretation. *New J. Phys.* **15**, 073022 (2013).
42. Tonomura, A., Endo, J., Matsuda, T. & Kawasaki, T. *Am. J. Phys.* **57**, 117–120 (1989).
43. Arlt, J. Handedness and azimuthal energy flow of optical vortex beams. *J. Mod. Opt.* **50**, 1573–1580 (2003).
44. Hamazaki, J., Mineta, Y., Oka, K. & Morita, R. Direct observation of Gouy phase shift in a propagating optical vortex. *Opt. Express* **14**, 8382–8392 (2006).
45. Cui, H.X., Wang, X.L., Gu, B., Li, Y.N., Chen, J. & Wang, H.T. Angular diffraction of an optical vortex induced by the Gouy phase. *J. Opt.* **14**, 055707 (2012).
46. Guzzinati, G., Schattschneider, P., Bliokh, K.Y., Nori, F. & Verbeeck, J. Observation of the Larmor and Gouy rotations with electron vortex beams. *Phys. Rev. Lett.* **110**, 093601 (2013).
47. Petersen, T.C., Paganin, D.M., Weyland, M., Simula, T.P., Eastwood, S.A. & Morgan, M.J. Measurement of the Gouy phase anomaly for electron waves. *Phys. Rev. A* **88**, 043803 (2013).
48. Aharonov Y. & Bohm, D. Significance of Electromagnetic Potential in the Quantum Theory. *Phys. Rev.* **115**, 485–491 (1959).

## Acknowledgements

This work was partially supported by the Austrian Science Fund (FWF) (grant No. I543-N20), RIKEN iTHES Project, MURI Center for Dynamic Magneto-Optics, JSPS-RFBR contract no. 12-02-92100, Grant-in-Aid for Scientific Research (S), MEXT Kakenhi on Quantum Cybernetics, and the JSPS via its FIRST program.